\documentclass[reprint,prl,aps]{revtex4-1}
\usepackage{graphicx}
\usepackage{color}
\usepackage{dcolumn}
\usepackage{bm}
\usepackage{amssymb}
\usepackage{color,amsmath}
\usepackage{soul,xcolor} %use \st{text} to achieve strikethrough
\setstcolor{red} %set colour of strikethrough to red
\usepackage{epstopdf}

\newcolumntype{C}[1]{>{\centering\arraybackslash}m{#1}}
\newcolumntype{N}{@{}m{0pt}@{}}

\definecolor{cadmiumgreen}{rgb}{0.0, 0.42, 0.24}

\begin{document}
% \linenumbers

\title{Tunable crystal symmetry in graphene--boron nitride heterostructures with coexisting moir\'e superlattices}

\author{Nathan R. Finney$^{1*}$} 
\author{Matthew Yankowitz$^{2*}$}
\author{Lithurshanaa Muraleetharan$^{1}$} 
\author{K. Watanabe$^{3}$} 
\author{T. Taniguchi$^{3}$} 
\author{Cory R. Dean$^{2\dagger}$}
\author{James Hone$^{1\dagger}$}  

\affiliation{$^{1}$Department of Mechanical Engineering, Columbia University, New York, NY, USA}
\affiliation{$^{2}$Department of Physics, Columbia University, New York, NY, USA}
\affiliation{$^{3}$National Institute for Materials Science, 1-1 Namiki, Tsukuba 305-0044, Japan}
\affiliation{$^{*}$These authors contributed equally to this work.}
\affiliation{$^{\dagger}$ cd2478@columbia.edu (C.R.D.); jh2228@columbia.edu (J.H.)}

%\begin{abstract}
%(abstract) 
%\end{abstract}

\maketitle

\textbf{
In heterostructures consisting of atomically thin crystals layered on top of one another, lattice mismatch or rotation between the layers results in long-wavelength moir\'e superlattices. These moir\'e patterns can drive significant band structure reconstruction of the composite material, leading to a wide range of emergent phenomena including superconductivity~\cite{Cao2018b,Yankowitz2019,Chen2019}, magnetism~\cite{Sharpe2019}, fractional Chern insulating states~\cite{Spanton2018}, and moir\'e excitons~\cite{Seyler2019,Tran2019,Jin2019,Alexeev2019}. Here, we investigate monolayer graphene encapsulated between two crystals of boron nitride (BN), where the rotational alignment between all three components can be varied. We find that band gaps in the graphene arising from perfect rotational alignment with both BN layers can be modified substantially depending on whether the relative orientation of the two BN layers is 0 or 60 degrees, suggesting a tunable transition between the absence or presence of inversion symmetry in the heterostructure. Small deviations ($<1^{\circ}$) from perfect alignment of all three layers leads to coexisting long-wavelength moir\'e potentials, resulting in a highly reconstructed graphene band structure featuring multiple secondary Dirac points. Our results demonstrate that the interplay between multiple moir\'e patterns can be utilized to controllably modify the electronic properties of the composite heterostructure.}

The ability to combine diverse vdW materials into a heterostructure enables engineering of new properties not observed in the constituent materials alone. A unique degree of freedom within these vdW heterostructures is the twist angle between layers, and changing this angle can strongly modify the material properties owing to the formation of moir\'e patterns. In graphene--BN heterostructures, the moir\'e pattern introduces a spatially-periodic effective potential that modifies the graphene band structure, giving rise to emergent secondary Dirac points (SDPs) at finite energy~\cite{Yankowitz2012,Ponomarenko2013,Dean2013,Hunt2013} and band gaps at the charge neutrality point and valence band SDP~\cite{Hunt2013,Woods2014,Chen2014,Wang2015,Wang2016,Song2013,Bokdam2014,Moon2014,Wallbank2015,Jung2015,SanJose2014,Slotman2015,Jung2017,Yankowitz2018}. In twisted bilayer graphene (tBLG), correlated insulating states and superconductivity emerge at a twist angle of $\sim$1.1$^{\circ}$ where the lowest energy moir\'e bands become exceptionally flat~\cite{Cao2018a,Cao2018b,Yankowitz2019,Sharpe2019,Lu2019}. However, typical vdW heterostructures comprising many flakes possess numerous crystal interfaces, and in principle multiple long-wavelength moir\'e patterns may coexist within a single heterostructure, likely with profound consequences on moir\'e-driven physics. For example, topological bands have been shown to potentially arise in tBLG aligned to BN~\cite{Sharpe2019}. So far, little has been done to controllably tune the alignment of multiple pairs of crystals within a single device, and it is not well understood how multiple moir\'e patterns interact to influence the properties of the vdW heterostructure.

%%%%%%%%%%%%%%%%%%%%%%%%%%%%%%%%%%%%%%%%%%%%%%%%%%%%%%%%%%%%%%%%%%%%%%
\begin{figure*}[ht]
\includegraphics[width=3.5 in]{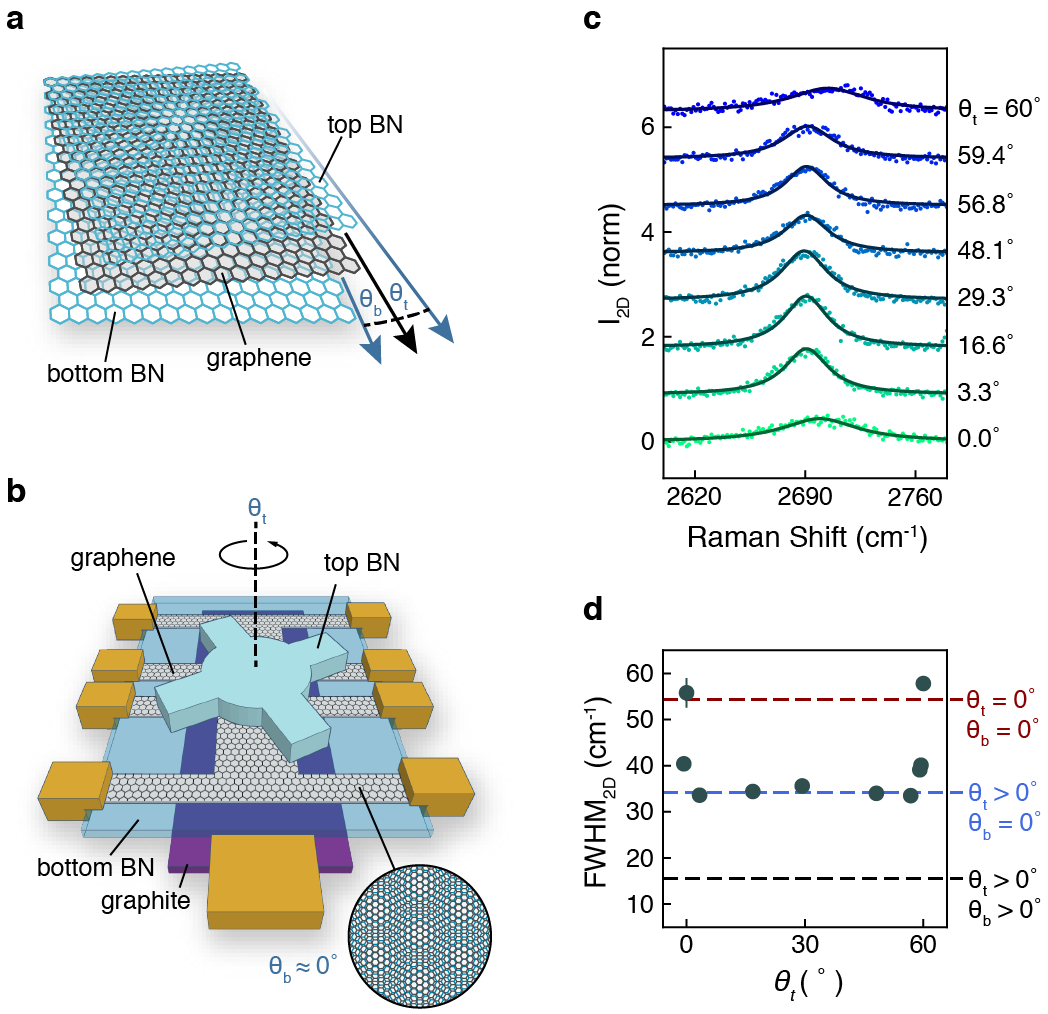} 
\caption{\textbf{Aligning top and bottom BN flakes to graphene.}
\textbf{a}, Schematic of graphene encapsulated in BN, with twist angles $\theta_b$ and $\theta_t$ between the layers.
\textbf{b}, Schematic of a graphene Hall bar device with a dynamically rotatable top BN flake. The graphene and bottom BN are perfectly aligned ($\theta_b\approx0^{\circ}$).
\textbf{c}, Graphene 2D Raman mode collected at several values of $\theta_t$ in a single rotatable device (R1) with $\theta_b\approx0^{\circ}$. Peaks are normalized to the intensity of the 2D peak at $\theta_t=29.3^{\circ}$
\textbf{d}, Full width at half maximum of the graphene 2D Raman peak as a function of $\theta_t$ (gray circles). Dashed lines denote FWHM$_{2D}$ for the conditions of graphene aligned to neither BN layer (black), to one BN layer (blue), and to both BN layers (red), taken from three stationary devices (see Supplementary S3).
}
\label{fig:1}
\end{figure*}
%%%%%%%%%%%%%%%%%%%%%%%%%%%%%%%%%%%%%%%%%%%%%%%%%%%%%%%%%%%%%%%%%%%%%%

In a heterostructure where graphene is encapsulated on both sides by BN, there are a number of qualitatively distinct stacking orders that can be realized by independently controlling the twist angle of the graphene relative to the bottom BN, $\theta_b$, and the top BN, $\theta_t$ (angles illustrated schematically in Fig.~\ref{fig:1}a). In this work, we fabricate devices in which the graphene is aligned to the bottom BN ($\theta_b=0^{\circ}\pm0.15^{\circ})$, but $\theta_t$ can be tuned to arbitrary angle. In this structure there are two unique positions in which the top BN layer may be ``aligned'' to the graphene --- $\theta_t=0^{\circ}$ and $60^{\circ}$ --- which have distinct symmetry due to the inequivalence of the boron and nitrogen triangular sublattices of the BN unit cell. Additionally, we study the case of  small misalignment of the top layer (\textit{i.e.} small but non-zero values of $\theta_t$) such that the two interfaces give rise to separate and incommensurate moir\'e patterns but with similar wavelengths.

%%%%%%%%%%%%%%%%%%%%%%%%%%%%%%%%%%%%%%%%%%%%%%%%%%%%%%%%%%%%%%%%%%%%%%
\begin{figure*}[ht]
\includegraphics[width=6.9 in]{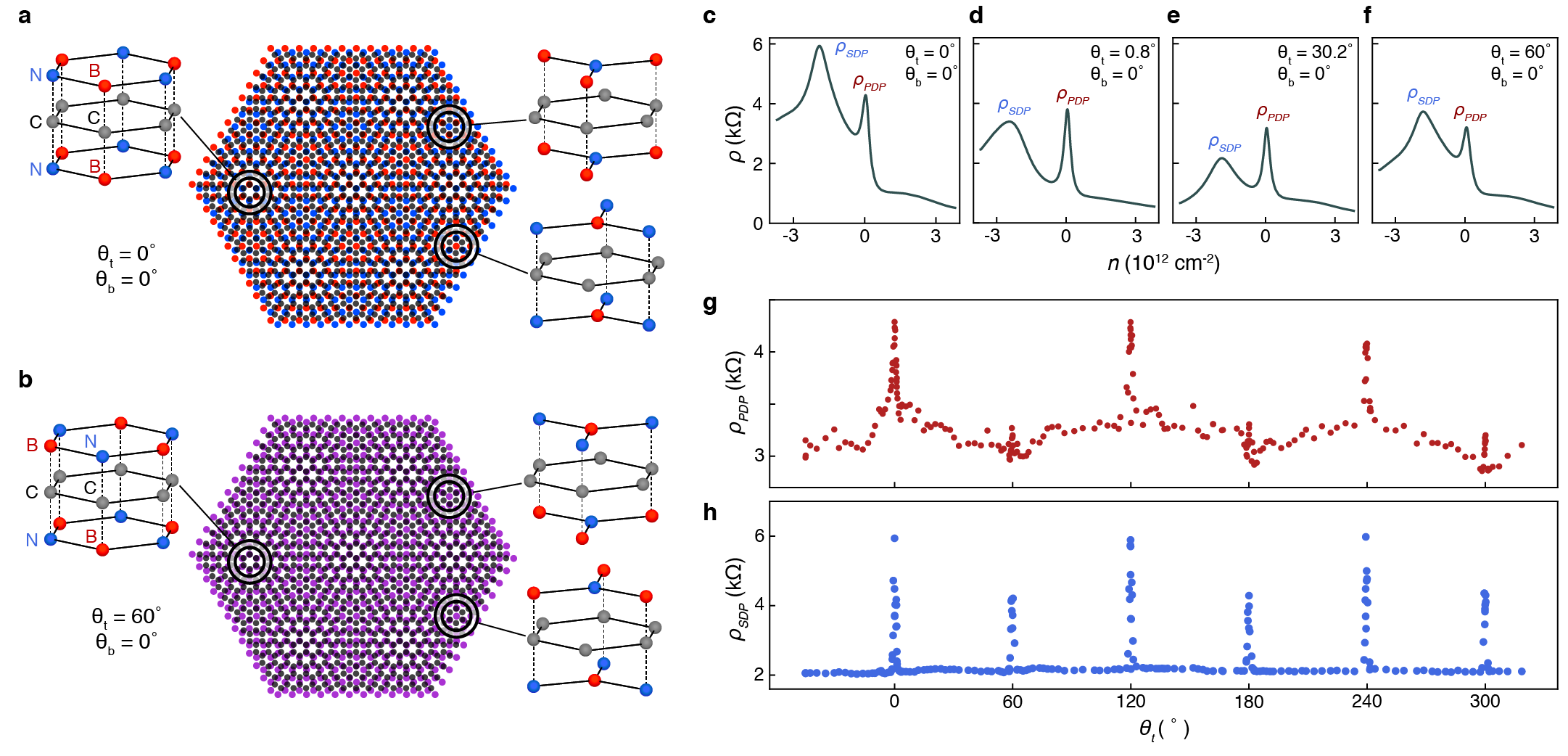} 
\caption{\textbf{Crystal symmetry and room temperature transport as a function of $\theta_t$.}
Schematics of BN--graphene--BN heterostructures with all layers aligned and \textbf{a,} 0$^{\circ}$ or \textbf{b}, 60$^{\circ}$ rotational offset between the top and bottom BN layers. Lattice models at the high symmetry points of the moir\'e pattern show the atomic alignment for each.
$\rho(n)$ at $T$ = 300 K for \textbf{c}, $\theta_t=0^{\circ}$, \textbf{d}, $\theta_t=0.8^{\circ}$, \textbf{e}, $\theta_t=30.2^{\circ}$, and \textbf{f}, $\theta_t=60^{\circ}$. 
\textbf{g}, $\rho_{PDP}$ and \textbf{h}, $\rho_{SDP}$ over a 360$^{\circ}$ range of $\theta_t$. 
}
\label{fig:2}
\end{figure*}
%%%%%%%%%%%%%%%%%%%%%%%%%%%%%%%%%%%%%%%%%%%%%%%%%%%%%%%%%%%%%%%%%%%%%%

As illustrated in Fig.~\ref{fig:1}b, we utilize a dynamically rotatable heterostructure to vary $\theta_t$ while maintaining fixed $\theta_b$. Graphene is first aligned to the bottom BN layer and shaped into a Hall bar geometry. The rotational alignment of the graphene and BN is initially determined by Raman spectroscopy prior to making electrical contact, and subsequently confirmed precisely by measuring the charge carrier density of the SDPs in electrical transport measurements (see Methods). The top BN is separately patterned into a circular shape with rectangular ``handles," and is subsequently transferred onto the graphene Hall bar such that it covers the entire active area of the channel (Fig.~\ref{fig:1}b). The top BN can be mechanically rotated to any angle ($0^{\circ}<\theta_{t}<360^{\circ}$) using an AFM tip, and electrical transport measurements can be performed simultaneously~\cite{Chari2016,Ribeiro2018}. We are able to measure changes in the twist angle of the top BN ($\Delta\theta_t$) to better than 0.1$^{\circ}$ from AFM topographs. 

Fig.~\ref{fig:1}c shows spectra of the graphene 2D Raman peak at different values of $\theta_t$. In Fig.~\ref{fig:1}d we plot the full width at half maximum of the 2D peak (FWHM$_{2D}$) versus $\theta_t$ over a 60$^{\circ}$ range. Over most of this range the FWHM$_{2D}$ exhibits a constant value that is $\sim$20 cm$^{-1}$ larger than isolated graphene. This behavior is consistent with the presence of a single long-wavelength moir\'e pattern resulting from the fixed zero-angle alignment between the graphene and bottom BN~\cite{Eckmann2013,Ribeiro2018}. When $\theta_t$ approaches $0^{\circ}$ or $60^{\circ}$, the linewidth is broadened by an additional $\sim$20 cm$^{-1}$, with FWHM$_{2D}$ near 55 cm$^{-1}$. We interpret this additional broadening to result from rotational alignment of the top BN layer to the graphene/bottom BN, and therefore provides an absolute measure of the top BN layer orientation. In previous studies, the broadening of the 2D mode was understood to arise from moir\'e-scale relaxations of the graphene lattice~\cite{Eckmann2013}. Our observation of an approximate doubling of this broadening indicates that the graphene couples similarly to the moir\'e patterns from both the top and bottom encapsulating BN layers.

While the 2D Raman peak is 60$^{\circ}$ periodic, the BN--graphene--BN trilayer lattice structure is not precisely equivalent upon 60$^{\circ}$ rotations of the top BN crystal. As illustrated schematically in Figs.~\ref{fig:2}a-b, for $\theta_t=0^{\circ}$ the moir\'e pattern is three-fold rotationally symmetric, whereas for $\theta_t=60^{\circ}$ the moir\'e pattern is six-fold rotationally symmetric. The lattice structures at the high-symmetry points of the moir\'e patterns (outer schematics in Figs.~\ref{fig:2}a-b) highlight the important difference between these two cases. For $\theta_t=0^{\circ} $ the top layer B (N) atoms sit atop the bottom layer B (N) atoms and the overall structure breaks inversion symmetry. In contrast, for $\theta_t=60^{\circ}$ the top layer B (N) atoms sit atop N (B) atoms and the structure hosts inversion symmetry. The nature of inversion symmetry in graphene--BN heterostructures has previously been tied to band structure reconstruction of the graphene~\cite{Hunt2013,Wang2015}, suggesting that the electronic properties of our devices may vary substantially with $\theta_t$.

We first investigate room temperature electrical transport of our device, and in particular observe notable differences between $\theta_t=0^{\circ}$ and 60$^{\circ}$. Figs.~\ref{fig:2}c-f show the device resistivity, $\rho$, as a function of charge carrier density, $n$, for various values of $\theta_t$. When the top BN is far from alignment (Fig. 2e), we observe both a sharp resistance peak at the primary Dirac point (PDP) and broad resistance peaks at finite density corresponding to the moir\'e-induced SDPs. We note that the resistivity of the hole-band SDP is larger than the electron-band SDP but smaller than the PDP, consistent with previous room temperature studies of graphene aligned to a single BN layer~\cite{Wang2015}. As $\theta_t$ approaches zero (Figs.~\ref{fig:2}c-d), both the PDP and SDP grow, but their relationship inverts with $\rho_{SDP}$ exceeding $\rho_{PDP}$ very near $\theta_{t}=0^{\circ}$. By contrast, near $\theta_{t}=60^{\circ}$ (Fig.~\ref{fig:2}f), the SDP again grows, but less so, and the PDP appears slightly suppressed. The angle dependence of the PDP and hole-band SDP peaks are shown in more detail in Figs.~\ref{fig:2}g-h, where the peak resistance values are plotted over a full $360^{\circ}$ rotation. In particular, we note that both the PDP and hole-band SDP are maximal at the angle we label $\theta_{t}=0^{\circ}$. While translation of the top BN with respect to the bottom would change the overall stacking configuration, we observe nearly equivalent transport every $\Delta\theta_t=120^{\circ}$. This suggests translation does not play a significant role, and that the stacking configuration with no translational offset is the structural ground state at the aligned positions.

In graphene aligned to a single BN layer, the staggered sublattice potential of the BN breaks inversion symmetry in the graphene layer for both 0$^{\circ}$ and 60$^{\circ}$ ``aligned'' orientations, resulting in a band gap at the PDP whose value is expected to scale with the magnitude of effective superlattice potential~\cite{Hunt2013,Woods2014,Chen2014,Wang2015,Wang2016,Song2013,Bokdam2014,Moon2014,Wallbank2015,Jung2015,SanJose2014,Slotman2015,Jung2017,Yankowitz2018}. We conjecture that the asymmetry between $\theta_{t}=0^\circ$ and 60$^\circ$ in our device correlates with the transition between the broken inversion symmetry structure at $\theta_{t}=\theta_{b}=0^\circ$ (Fig.~\ref{fig:2}a) --- in which the PDP gap is likely to be largest --- and the inversion symmetric structure at $\theta_{t}=60^\circ$ (Fig.~\ref{fig:2}b) --- in which no PDP gap would be expected within a single-particle model. 

%%%%%%%%%%%%%%%%%%%%%%%%%%%%%%%%%%%%%%%%%%%%%%%%%%%%%%%%%%%%%%%%%%%%%%
\begin{figure*}[ht]
\includegraphics[width=5.5 in]{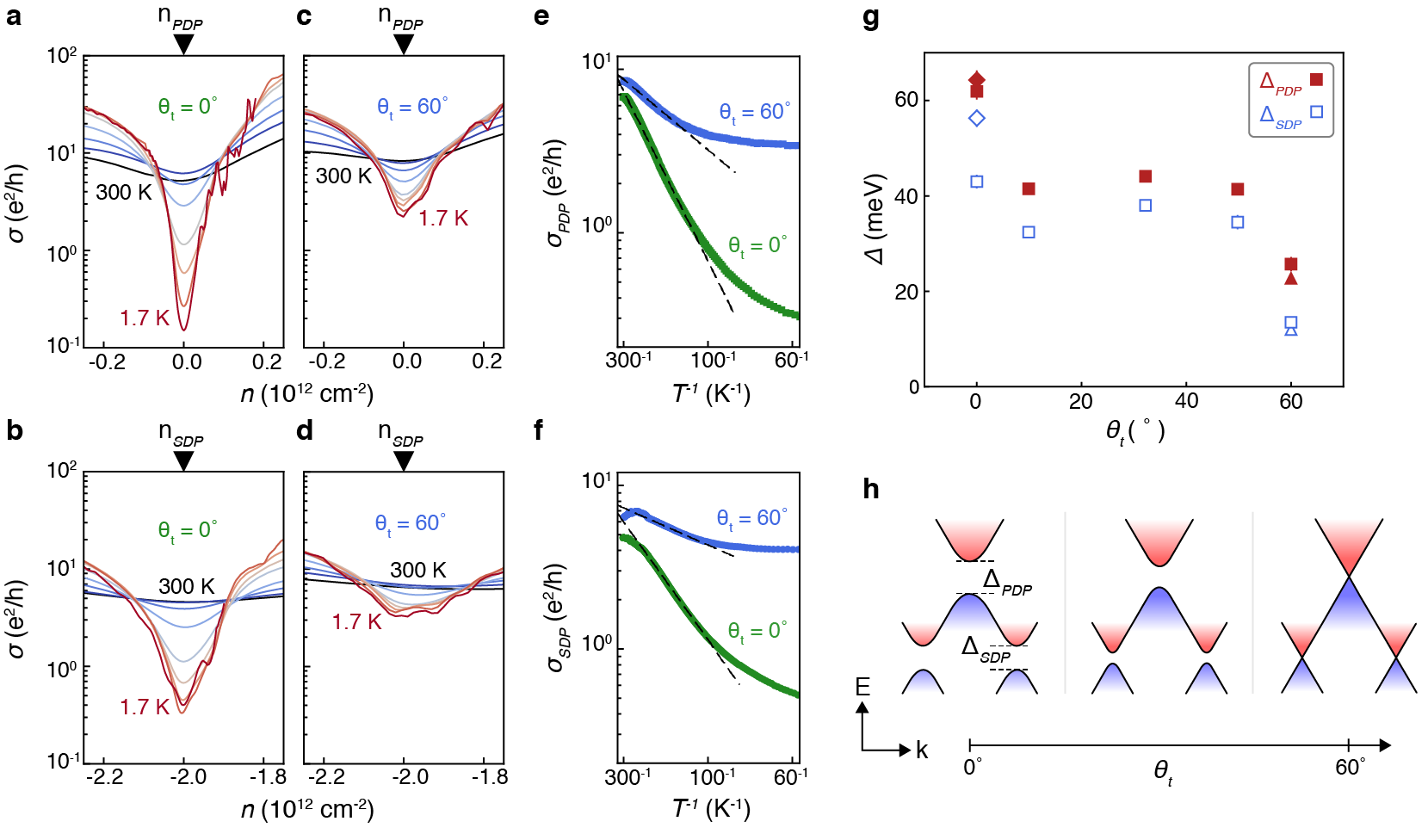} 
\caption{\textbf{Graphene band gaps as a function of $\theta_t$ in devices with $\theta_b\approx0^{\circ}$.}
Conductivity of device R1 at several temperatures near the \textbf{a,} PDP and \textbf{b,} SDP for $\theta_t=0^{\circ}$. 
Same at the \textbf{c,} PDP and \textbf{d,} SDP for $\theta_t=60^{\circ}$.
Arrhenius plot of the conductivity of the \textbf{e,} PDP and \textbf{f,} SDP as a function of inverse temperature at $\theta_t=0^{\circ}$ and $\theta_t=60^{\circ}$. The slope of the linear fits (black dashed lines) give the band gap $\Delta_{p,s}$ via $\sigma_{p, s}(T) \propto e^{-\frac{\Delta}{2 k T}}$, where $k$ is the Boltzmann constant.
\textbf{g}, $\Delta_{PDP}$ (closed red markers) and $\Delta_{SDP}$ (open blue markers) for rotatable device R1 (square markers), and for stationary devices S1 (diamond markers) and S2 (triangle markers). The error bars in the gaps are set by the determination of the linear (thermally activated) regime for the fit. The uncertainty in determining $\theta_t$ is smaller than the width of the markers.
\textbf{h}, Cartoon illustration of the graphene band structure as a function of $\theta_t$. Both gaps are largest for $\theta_t=0^{\circ}$, while inversion symmetry protects the Dirac crossings at $\theta_t=60^{\circ}$.
}
\label{fig:3}
\end{figure*}
%%%%%%%%%%%%%%%%%%%%%%%%%%%%%%%%%%%%%%%%%%%%%%%%%%%%%%%%%%%%%%%%%%%%%%

Figs.~\ref{fig:3}a-d compare the temperature dependence of the PDP and SDP at $\theta_t=0^{\circ}$ and 60$^{\circ}$. Strongly insulating behavior is observed in both at $\theta_t=0^{\circ}$, whereas only weakly insulating behavior is observed for $\theta_t=60^{\circ}$. A linear fit to the thermally activated regime (black dashed lines in Figs.~\ref{fig:3}e-f) gives a measure of the activation gap, $\Delta$. In Fig.~\ref{fig:3}g we plot the gaps for 5 different values of $\theta_t$ (square markers). Both the PDP and SDP gaps are notably enhanced at $\theta_t=0^{\circ}$, whereas at $\theta_t=60^{\circ}$ both are notably reduced. The gaps have little dependence on $\theta_t$ at all other angles. Our observation of $\Delta_{PDP}>60$ meV is so far the largest gap observed in a pristine graphene device, and may potentially be significantly further enhanced under pressure~\cite{Yankowitz2018}. Although the gaps extracted from an Arrhenius fit are not identically zero at $\theta_t=60^{\circ}$, the device only exhibits activated behavior over well less than a decade change in conductance, hence we expect our reported gaps to be an upper bound at this angle. 

The band structure modification anticipated from symmetry considerations is illustrated schematically in Fig.~\ref{fig:3}h, where we anticipate the largest gaps for $\theta_t=\theta_b=0^{\circ}$ owing to the doubled moir\'e potential, while at $\theta_t=60^{\circ}$ inversion symmetry protects the Dirac crossings. Following this simple expectation, the measured gaps corroborate our previous assignments of $\theta_t=0^{\circ}$ and $60^{\circ}$. We additionally measure the band gaps in two ``stationary'' devices (\textit{i.e.} without the ability to dynamically rotate the top BN, see Supplementary Section 2), which also exhibit large broadening of the FWHM$_{2D}$ Raman peak. The gaps are similarly either enhanced or suppressed (diamond and triangle markers in Fig.~\ref{fig:3}c), suggesting this effect is generic for samples in which both BN layers are aligned to graphene.

%%%%%%%%%%%%%%%%%%%%%%%%%%%%%%%%%%%%%%%%%%%%%%%%%%%%%%%%%%%%%%%%%%%%%%
\begin{figure*}[ht]
\includegraphics[width=6.3 in]{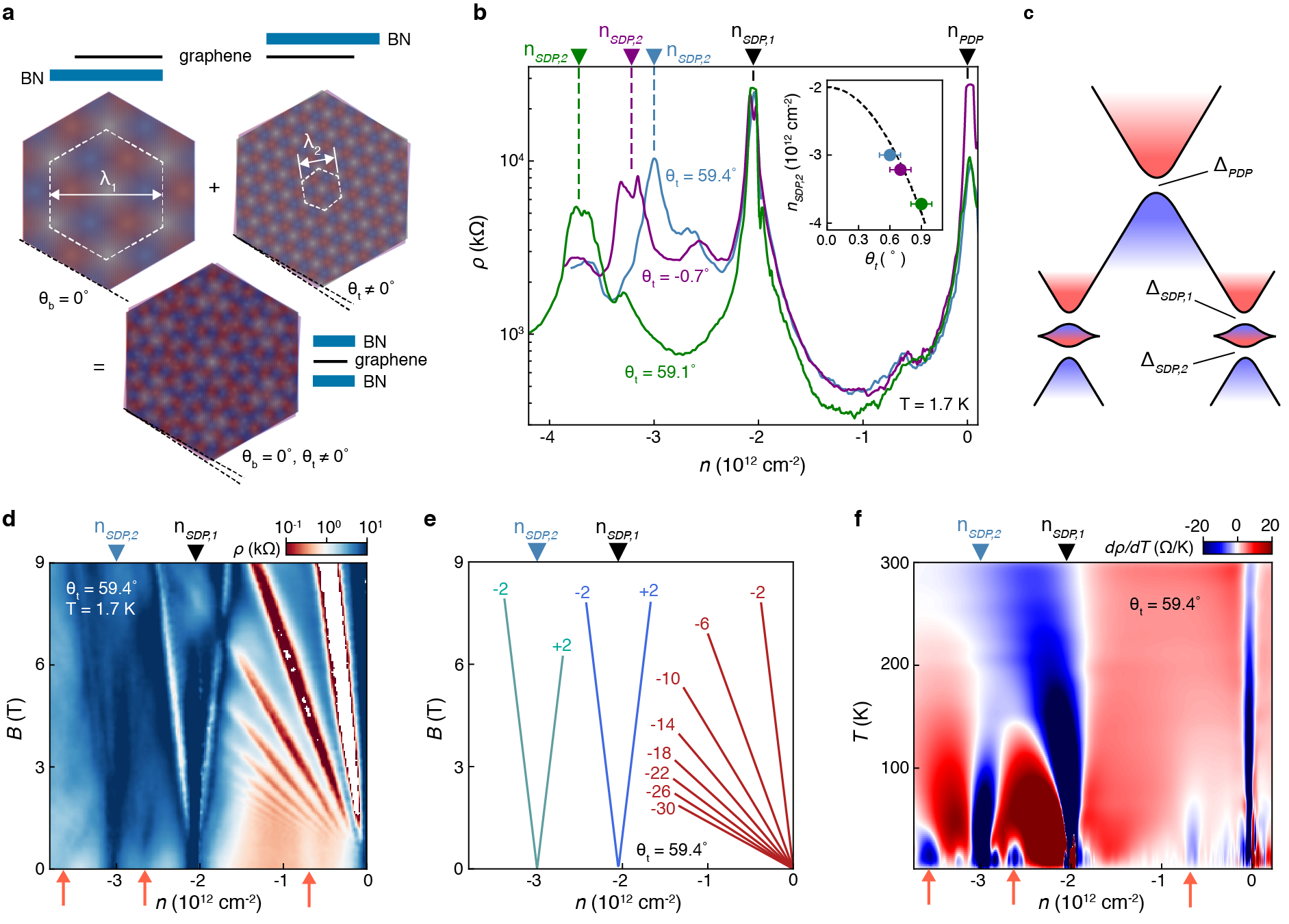} 
\caption{\textbf{Coexisting moir\'e structures in BN--graphene--BN heterostructures.}
\textbf{a}, Moir\'e pattern from graphene on BN ($\theta_b=0^{\circ}$) (top left) and BN on graphene ($\theta_t\neq0^{\circ}$) (top right). A linear combination of these moir\'e patterns in a BN-encapsulated graphene heterostructure (bottom middle) yields two coexisting long-wavelength moir\'e patterns, as well as an additional second-order moir\'e pattern arising from their interference.
\textbf{b}, $\rho(n)$ at $T$ = 1.7 K for device R1 with $\theta_t<1^{\circ}$ from perfect alignment. Triangle markers indicate the carrier density corresponding to the PDP (n$_{PDP}$), the SDP from the graphene and bottom BN (n$_{SDP,1}$), and the SDP from the graphene and top BN (n$_{SDP,2}$). (inset) $n_{SDP,2}$ versus $\theta_t$ measured from AFM topographs. Black dashed line shows the anticipated dispersion taking a lattice mismatch of $\delta \approx 1/60$.
\textbf{c}, Cartoon illustration of the graphene band structure exhibiting two different SDP gaps in the valence band.
\textbf{d}, Landau fan diagram with $\theta_t=59.4^{\circ}$ at $T$ = 1.7 K. Quantum oscillations with dominant sequence of $\nu=-2,-6,-10,...$ emerge from the PDP, while quantum oscillations of $\nu=\pm2$ emerge from both $n_{SDP,1}$ and $n_{SDP,2}$. Orange arrows denote weak signatures of additional states.
\textbf{e}, Schematic Landau level structure corresponding to the observations in (d).
\textbf{f}, $d\rho/dT$ as a function of $T$ and $n$. Activated behavior is observed at $n=0$, $n_{SDP,1}$ and $n_{SDP,2}$. Orange arrows indicate additional values of $n$ which exhibit negative (but not activated) $d\rho/dT$, and correspond to those in (d).
}
\label{fig:4}
\end{figure*}
%%%%%%%%%%%%%%%%%%%%%%%%%%%%%%%%%%%%%%%%%%%%%%%%%%%%%%%%%%%%%%%%%%%%%%

Finally, we investigate the small angle limit ($0^{\circ}<\theta_t<1^{\circ}$) where the top and bottom BN yield moir\'e patterns with only slightly different period. Fig.~\ref{fig:4}b shows the hole-doped transport for three values of $\theta_t\neq0^{\circ}$ in this regime. In addition to a peak in the resistivity at the PDP, we observe two sizable peaks at finite hole-doped densities denoted $n_{SDP,1}$ and $n_{SDP,2}$. $n_{SDP,1}$ is independent of $\theta_t$ and corresponds to the SDP arising from the moir\'e potential of the perfectly aligned graphene and bottom BN ($\theta_b\approx0^{\circ}$), while the resistance peaks at $n_{SDP,2}$ correspond to the moir\'e potential from the top BN. The position of $n_{SDP,2}$ as a function of $\theta_t$ is shown the inset of Fig.~\ref{fig:4}b, and is in good quantitative agreement with the anticipated dispersion (see Supplementary Section 4)~\cite{Yankowitz2012}. We note that in the small angle limit, thermally activated behavior is observed at both $n_{SDP,1}$ and $n_{SDP,2}$, suggesting band gaps associated with each SDP with typical values of $\sim$20 meV and $\sim$5 meV, respectively (see Supplementary Fig. 5). This implies the reconstructed band structure shown schematically in Fig.~\ref{fig:4}c in which an isolated narrow band emerges between the two superlattice gaps $\Delta_{SDP,1}$ and $\Delta_{SDP,2}$. Further theoretical and experimental effort is necessary to fully explore the consequence of this band reconstruction, however the appearance of a flat band whose width varies with rotation angle provides the intriguing possibility of hosting tunable correlated states at low temperature~\cite{Cao2018a,Chen2019}. 

In a magnetic field, we observe sequences of quantum oscillations emerging from the Dirac points at $n=0$, $n_{SDP,1}$, and $n_{SDP,2}$ (Fig.~\ref{fig:4}d-e, for $\theta_t=59.4^{\circ}$). We further observe weak signatures of resistive states adjacent to each of the three Dirac points, marked by orange arrows. To identify these resistive states more clearly, we plot $d\rho/dT$ versus $T$ and $n$ in Fig.~\ref{fig:4}f. In addition to the insulating states previously discussed at the PDP and the two SDPs, we observe a negative temperature dependence at the same densities marked by orange arrows in Fig.~\ref{fig:4}d, suggesting the presence of new insulating-like states. The two coexisting moir\'e patterns may in principle interfere to produce a second-order moir\'e pattern with a very long period (Fig.~\ref{fig:4}a), inducing an additional resistive state at low density~\cite{Wang2019}. However, the position of the low density resistive feature does not agree precisely with quantitative theoretical estimates for all measured $\theta_t$ (see Supplementary Section 6). The disagreement may arise due to unexpected structural reconstructions of the second order moir\'e pattern, however at this point we are unable to understand these new resistive states in detail.

In conclusion, we demonstrate the ability to induce and control multiple moir\'e patterns within a BN--graphene--BN heterostructure. In particular, we are able to dynamically tune the crystal symmetry of the composite material by realizing distinct stacking configurations of the three layers, and further induce coexisting moir\'e patterns which combine to strongly modify the graphene band structure. Our techniques for engineering multiple moir\'e patterns within a single vdW heterostructure are quite general and can easily be extended to a wide variety of other device structures, motivating a new class of experiments in which the twist angle of multiple crystal interfaces can be tuned to realize novel material properties.

\section*{Methods}
All heterostructures are assembled using standard dry-transfer techniques with a poly-propylene carbonate (PPC) film on a polydimethyl siloxane (PDMS) stamp~\cite{Wang2013}, and rest atop a Si/SiO$_2$ substrate. The device fabrication of the rotatable devices largely follows the techniques developed in Refs.~\cite{Chari2016,Ribeiro2018}. Device R1 consists of a graphene Hall bar on a $\sim$44 nm thick BN resting on a $\sim$11 nm thick graphite gate. A $\sim$56 nm thick BN flake capped by $\sim$40 nm of hydrogen silsesquioxane (HSQ) is subsequently transferred onto the graphene Hall bar. The HSQ cap acts as the etch mask to shape the rotating BN, and also provides additional durability during AFM pushes. ``Stationary'' devices consist of a graphene Hall bar fully encapsulated by BN, all atop a graphite gate. Electrical contact to all devices is made by Cr/Au (2 nm/100 nm). Supplementary Sections~1-2 and Supplementary Figs.~1-2 provide full details of the device fabrication.

Raman spectroscopy measurements are performed at room temperature in air. All measurements are acquired using a 532 nm wavelength laser with a power $<$ 1 mW. Measurements to extract the graphene/BN twist angle in stationary devices are performed before the addition of the graphite back gate. In the rotatable device, the encapsulated graphene region sits atop the graphite gate. To isolate the Raman response from the graphene, we separately acquire a Raman spectrum from a nearby region of the graphite gate without the graphene and subtract this background response. We extract FWHM$_{2D}$ from a Lorentzian fit (see Supplementary Section 3 and Supplementary Fig. 3 for full details).

Transport measurements are conducted in a four-terminal geometry with ac current excitation of 10-100 nA using standard lock-in technique at 17.7 Hz. The graphene contact regions (which extend beyond the graphite bottom gate) are gated to a high carrier density by applying a bias to the silicon substrate (typically 5-50 V for SiO$_2$ thickness of $\sim$285 nm) to reduce the contact resistance. We extract $n(V_G)$ by fitting the dispersion of the quantum Hall states in high magnetic field as $n=\nu eB/h$, where $\nu$ is the filling factor, $h$ is Planck's constant, and $e$ is the elementary charge. The moir\'e wavelength, $\lambda$, is calculated using the geometric relation $\lambda^2 = 8/(n_{SDP}\sqrt{3})$ where n$_{SDP}$ is the density at full filling of the moir\'e unit cell.

\section*{acknowledgments}
The authors thank Rebeca Ribeiro-Palau, Changjian Zhang, and Shaowen Chen for technical support, as well as Jeil Jung, Mikito Koshino and Chris Marianetti for helpful discussions. This work was primarily supported by the NSF MRSEC program through Columbia in the Center for Precision Assembly of Superstratic and Superatomic Solids (DMR-1420634). Sample device design and fabrication was partially supported by DoE Pro-QM EFRC (DE-SC0019443). N.F. acknowledges support from the Stewardship Science Graduate Fellowship program provided under cooperative agreement number DE-NA0002135. CRD acknowledges the support of the David and Lucile Packard Foundation. K.W. and T.T. acknowledge support from the Elemental Strategy Initiative conducted by the MEXT, Japan and  the CREST (JPMJCR15F3), JST.

\section*{Author contributions}
N.R.F. and L.M. fabricated the devices. N.R.F. and M.Y. performed the measurements and analyzed the data. K.W. and T.T. grew the hBN crystals. C.R.D. and J.H. advised on the experiments. The manuscript was written with input from all authors.

\section*{Competing interests}
The authors declare no competing interests.

\bibliographystyle{naturemag}
\bibliography{references}

\clearpage

%%%%%%%%%%%%%%%%%%%%%%%%%%%%%%%%%%%%%%%%%%%%%%%%%%%%%%%%%%%%%%%%%%%%%%%%%%%%%%%%%%%%%%%

\renewcommand{\thefigure}{S\arabic{figure}}
\renewcommand{\thesubsection}{S\arabic{subsection}}
\setcounter{secnumdepth}{2}
\renewcommand{\theequation}{S\arabic{equation}}
\renewcommand{\thetable}{S\arabic{table}}
%\subsubsectionfont{\normalfont\large\itshape\underline}
\setcounter{figure}{0} 
\setcounter{equation}{0}

\section*{Supplementary Information}

%%%%%%%%%%%%%%%%%%%%%%%%%%%%%%%%%%%%%%%%%%%%%%%%%%%%%%%%%%%%%%%%%%%%%%
\begin{figure*}[ht]
\includegraphics[width=7.0 in]{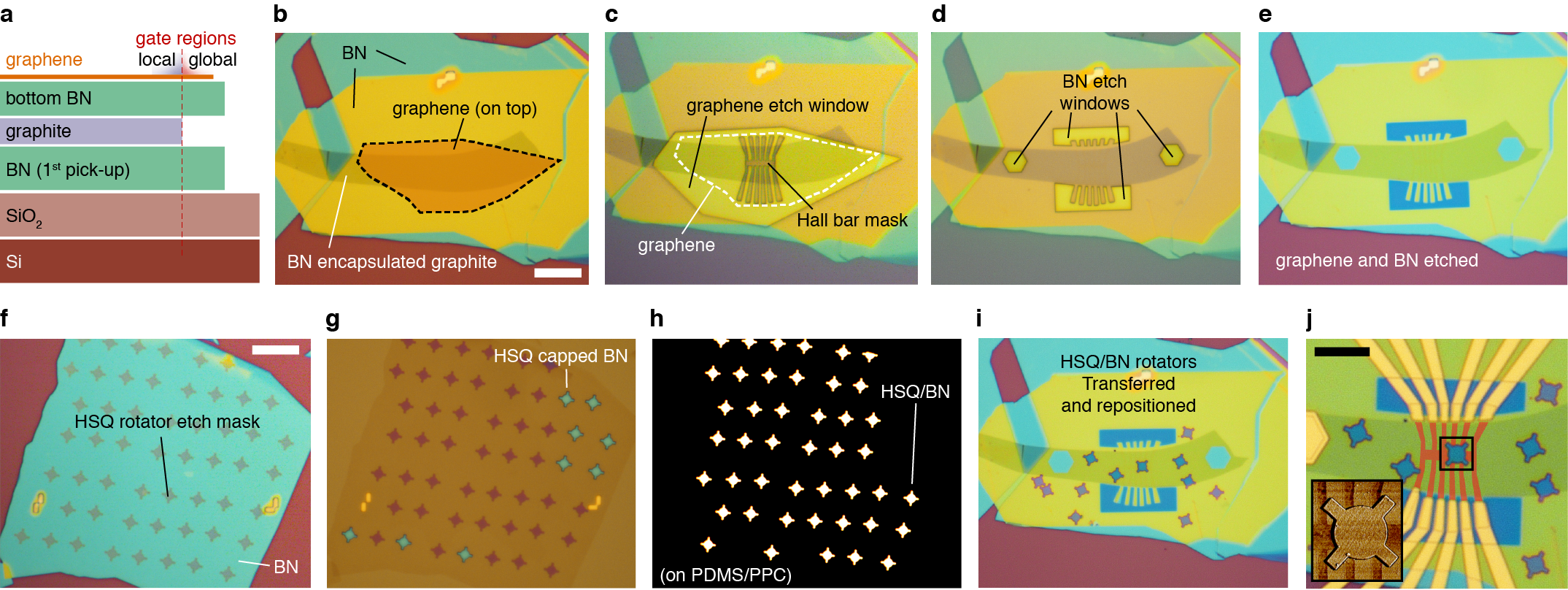} 
\caption{\textbf{Fabrication of rotatable devices.}
\textbf{a,} Schematic side-view of the bottom portion of the heterostructure. The graphene Hall bar channel is locally gated by graphite, and the graphene contact regions are gated by the silicon.
\textbf{b,} Optical micrograph of the heterostructure shown in \textbf{a} (top view), with false coloring applied to the graphene (highlighted in orange). Scale bar is 10 $\mu$m.
\textbf{c,} Graphene Hall bar etch mask defined in PMMA. The outline of graphene prior to etching is shown with a white dashed line.
\textbf{d,} PMMA etch mask to shape the BN.
\textbf{e,} Processed heterostructure after sequentially etching the graphene and BN using the masks in \textbf{c} and \textbf{d,} respectively.
\textbf{f,} HSQ masks on BN prior to etching. BN rotators are processed on a separate substrate from the heterostructure shown in \textbf{a}. The scale bar is 10 $\mu$m.
\textbf{g,} Etched HSQ-capped BN rotators (light blue in color), after picking up with PPC/PDMS dry stamp. The purple rotator shapes are pristine SiO$_2$ protected during etching of the BN, now exposed after picking up the HSQ/BN rotators.
\textbf{h,} HSQ/BN rotators on PPC/PDMS after pick-up from the substrate shown in \textbf{f} and \textbf{g}. The surrounding dark region is the PPC surface on the transfer stamp.
\textbf{i,} HSQ/BN rotators on processed heterostructure, after repositioning the rotators with an atomic force microscope (AFM) tip in contact mode, and cleaning with vacuum annealing at 350 $^{\circ}$C.
\textbf{j,} Fully processed device with metal contacts (100 nm gold with a 2 nm chromium adhesion layer), and false coloring applied to the graphene hall bar (shown in orange). The scale bar is 5 $\mu$m. The inset shows an AFM scan of the rotator over the channel region.
}
\label{fig:S1}
\end{figure*}
%%%%%%%%%%%%%%%%%%%%%%%%%%%%%%%%%%%%%%%%%%%%%%%%%%%%%%%%%%%%%%%%%%%%%%

%%%%%%%%%%%%%%%%%%%%%%%%%%%%%%%%%%%%%%%%%%%%%%%%%%%%%%%%%%%%%%%%%%%%%%
\begin{figure*}[ht]
\includegraphics[width=7.0 in]{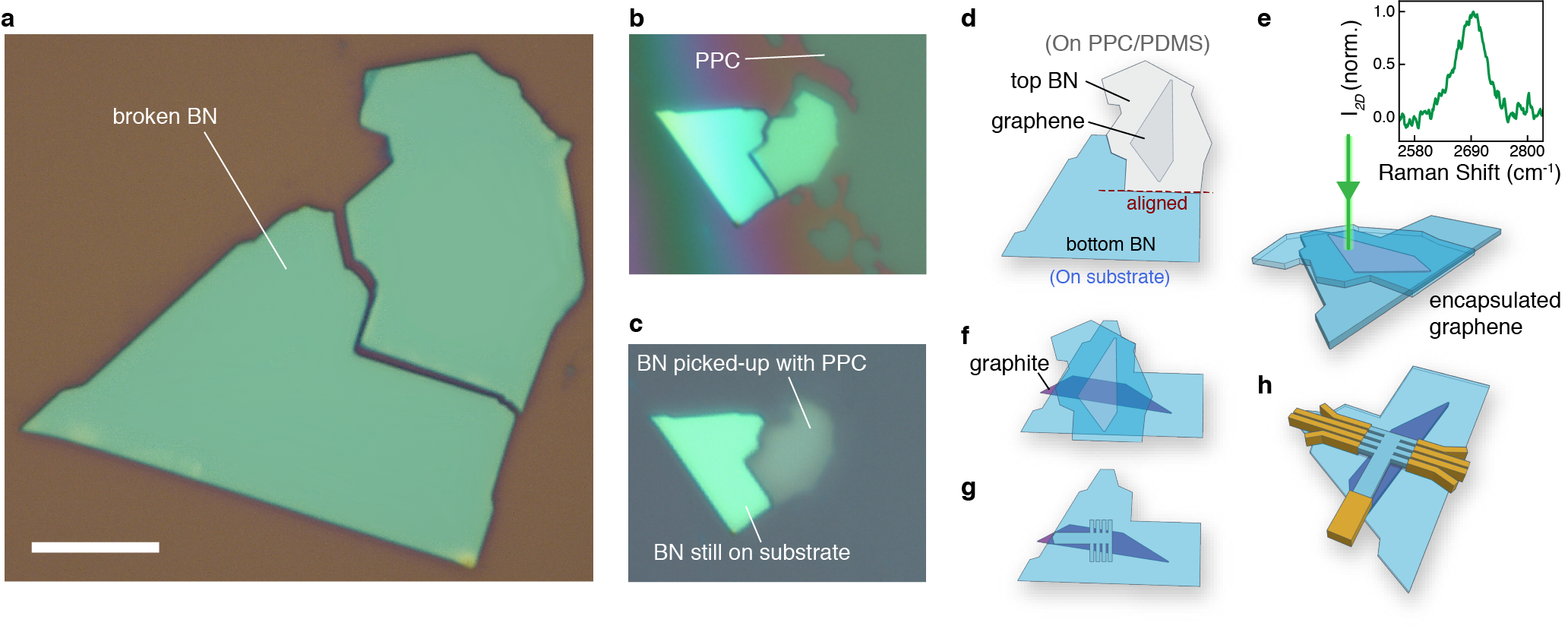}
\caption{\textbf{Fabrication of stationary devices.}
\textbf{a,} Optical micrograph image of a broken BN flake after mechanical exfoliation. Scale bar is 15 $\mu$m.
\textbf{b,} PPC covers the top section of the broken BN piece during the pick-up process.
\textbf{c,} After pick-up the top section of broken BN is located above the focal plane of the optical microscope, away from the substrate.
\textbf{d,} Schematic showing alignment of BN straight edges after picking up a piece of graphene with the top section of broken BN.
\textbf{e,} Schematic illustrating Raman characterization of the BN--graphene--BN heterostructure.
\textbf{f,} The heterostructure is picked up again, along with a graphite flake which acts as a back gate. It is deposited on a clean Si/SiO$_2$ substrate.
\textbf{g,} The graphene is shaped into a Hall bar, with portions of the contacts overhanging the graphite back gate.
\textbf{h,} Schematic of the completed stationary device after metal electrode deposition. 
}
\label{fig:S2}
\end{figure*}
%%%%%%%%%%%%%%%%%%%%%%%%%%%%%%%%%%%%%%%%%%%%%%%%%%%%%%%%%%%%%%%%%%%%%%

\subsection{Fabrication of rotatable devices}

To fabricate rotatable devices R1 and R2, we begin by sequentially picking up flakes of BN, graphite, BN, and graphene using a PPC/PDMS stamp. The PPC film is then mechanically removed from the PDMS stamp and placed onto a Si/SiO$_2$ substrate, such that the graphene layer is exposed at the top of the heterostructure. The underlying PPC is removed by vacuum annealing at $T=350^{\circ}$ C. In all processed devices the silicon is used as a global gate for the graphene contacts, and the graphite is used as a local gate for the graphene channel (Fig.~\ref{fig:S1}a-b). The use of a graphite gates has previously been demonstrated to significantly improve the charge homogeneity of graphene devices~\cite{Zibrov2017}.

Devices with good rotational alignment of the graphene and BN are next identified using Raman spectroscopy~\cite{Eckmann2013,Ribeiro2018}. Devices in which the graphene and BN are well aligned are then shaped into a Hall bar geometry using an oxygen plasma etch through a PMMA mask defined by standard electron-beam lithography (Fig.~\ref{fig:S1}c). In order to realize more robust electrical contacts, the BN surrounding the graphene contacts is further etched using a CHF$_3$/O$_2$ plasma through a PMMA mask (Fig.~\ref{fig:S1}d). All electrical contacts in device R1 show no significant change over 8 thermal cycles ($T$ = 300 K - 1.5 K), while many of the contacts in device R2 (in which no BN etch is performed) became open after thermal cycling.

We exfoliate BN onto a separate Si/SiO$_2$ substrate and etch an array of BN ``rotators'' using a protective HSQ mask (Figs.~\ref{fig:S1}f-g). The rotators are picked up using a separate PPC/PDMS stamp (Fig.~\ref{fig:S1}h) and transferred onto the graphene Hall bar (Fig.~\ref{fig:S1}i). A single rotator is aligned as desired onto the graphene Hall bar by positioning with an AFM tip in contact mode. The device is annealed a second time before deposition of Cr/Au (2 nm/100 nm) metal electrodes (Fig.~\ref{fig:S1}j). The graphene channel is further cleaned by sequentially pushing a BN rotator across the surface to collect and remove interfacial contamination.

\subsection{Fabrication of stationary devices}

To fabricate stationary devices S1-S4, we begin by identifying flakes of BN which have fractured into two pieces during the mechanical exfoliation procedure (Fig.~\ref{fig:S2}a). These fractured BN pieces share common edge profiles which can be easily optically aligned. One piece of the fractured BN is first picked up by a PPC/PDMS stamp (Fig.~\ref{fig:S2}b-c), leaving the remaining piece on the Si/SiO$_2$ wafer. A flake of graphene is then added to the heterostructure, and is chosen and aligned to fit completely within the BN area (Fig.~\ref{fig:S2}d). Next, the two BN flakes are rotationally aligned optically (with precision better than 0.5$^{\circ}$) and the remaining BN is added to the heterostructure (Fig.~\ref{fig:S2}e). The heterostructure is transferred to a Si/SiO$_2$ substrate and annealed at $T=350^{\circ}$ C. No effort is made to align crystal edges of graphene to the BN, however, because the graphene rests entirely within both BN flakes it tends to rotate to alignment with the BN during the transfer or annealing~\cite{Wang2015}. Devices with good rotational alignment of the graphene and BN are identified with Raman spectroscopy. A new PPC/PDMS stamp is then used to pick up the heterostructure, followed by a graphite flake (Fig.~\ref{fig:S2}f), and the heterostructure is transferred to a separate Si/SiO$_2$ substrate. The device is then shaped into a Hall bar geometry and electrically contacted following the procedures described for devices R1 and R2 (Figs.~\ref{fig:S2}g-h).

%%%%%%%%%%%%%%%%%%%%%%%%%%%%%%%%%%%%%%%%%%%%%%%%%%%%%%%%%%%%%%%%%%%%%%
\begin{figure*}[ht]
\includegraphics[width=6.5 in]{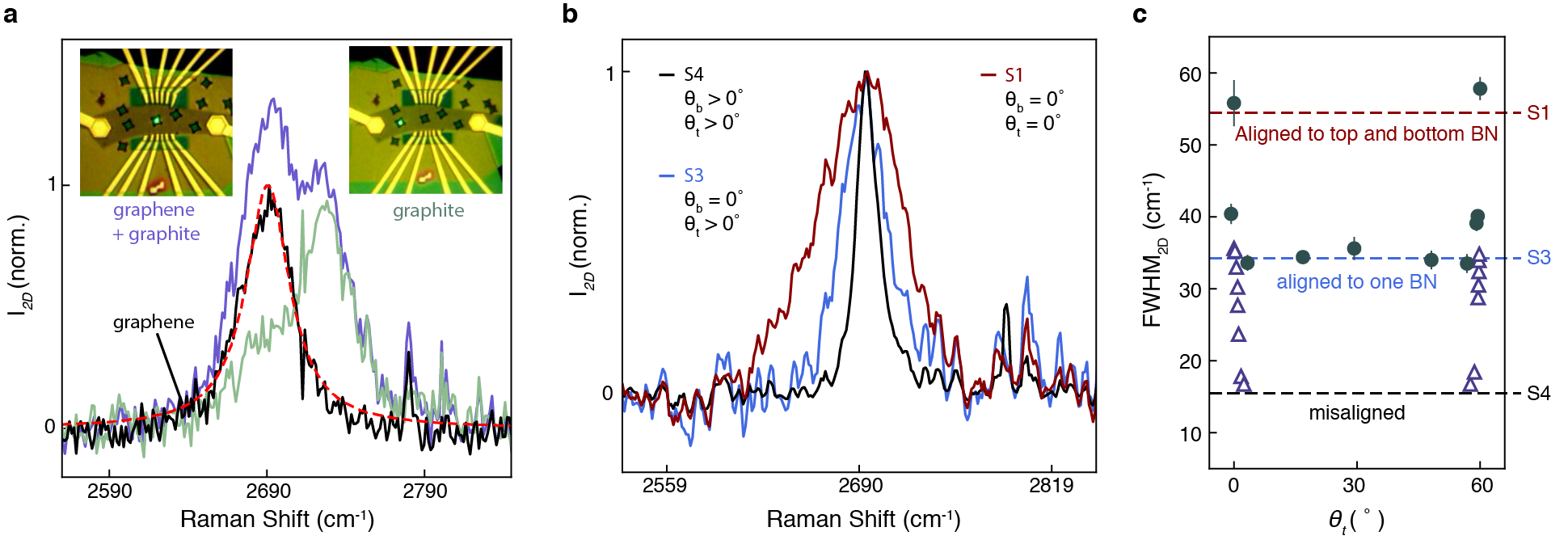} 
\caption{\textbf{Graphene 2D Raman peak collected for stationary and rotatable devices.}
\textbf{a,} Peak subtraction procedure utilized to isolate the graphene spectrum in rotatable devices with a graphite back gate. Insets show the position of the laser spot (green dot) in measurements to acquire the graphene+graphite (left) and graphite spectra (right), with corresponding spectra shown in purple and green, respectively. The difference of the purple and green curves is shown in black and taken to be the isolated response of the graphene. The red dashed line representing a single Lorentzian fit to the black curve.
\textbf{b,} Graphene 2D Raman peak for stationary BN encapsulated graphene devices in three different stacking configurations, including perfect alignment between all three layers (red curve), graphene aligned to just one of the two BN encapsulating layers (blue curve), and graphene aligned to neither encapsulating BN layer (black curve).
\textbf{c,} FWHM$_{2D}$ for several devices, including stationary devices S1, S3, and S4 (this work, horizontal dashed lines) and rotatable devices R1 (this work, filled dark gray circle markers) and a rotatable device with misaligned graphene and bottom BN from Ref.~\cite{Ribeiro2018} (open blue triangle markers).
}
\label{fig:S3}
\end{figure*}
%%%%%%%%%%%%%%%%%%%%%%%%%%%%%%%%%%%%%%%%%%%%%%%%%%%%%%%%%%%%%%%%%%%%%%

\subsection{Analysis of Raman spectroscopy}

Fig.~\ref{fig:S3}a plots Raman spectra near the 2D graphene peak in device R1. The measured signal (purple curve) features contributions from both the graphene channel and the graphite back gate. To isolate the graphene response (black curve), we separately acquire a Raman spectrum on the graphite gate alone (green curve) and subtract that contribution. The insets show the position of the laser (green dot) during acquisitions of the purple and green curves. We fit the graphene 2D peak with a Lorentzian (red dashed curve) to extract FWHM$_{2D}$. The reported values of FWHM$_{2D}$ are averaged across a few (3-6) measured background spots. Fig.~\ref{fig:S3}b shows the 2D Raman peak for stationary BN encapsulated graphene devices in three distinct stacking configurations. These spectra are acquired prior to the addition of graphite back gates.

Fig.~\ref{fig:S3}c plots the graphene FWHM$_{2D}$ at several alignment configurations for two different rotatable devices. The open triangle markers are taken from Ref.~\cite{Ribeiro2018}, in which the graphene is misaligned from the bottom BN. The closed circle markers reproduce the data from device R1 shown in Fig.~\ref{fig:1}d of the main text. The dashed lines show FWHM$_{2D}$ of stationary devices S1 (graphene aligned to both BN layers), S3 (graphene aligned to a single BN layer), and S4 (graphene not aligned to either BN layer). 

%%%%%%%%%%%%%%%%%%%%%%%%%%%%%%%%%%%%%%%%%%%%%%%%%%%%%%%%%%%%%%%%%%%%%%
\begin{figure*}[ht]
\includegraphics[width=7.0 in]{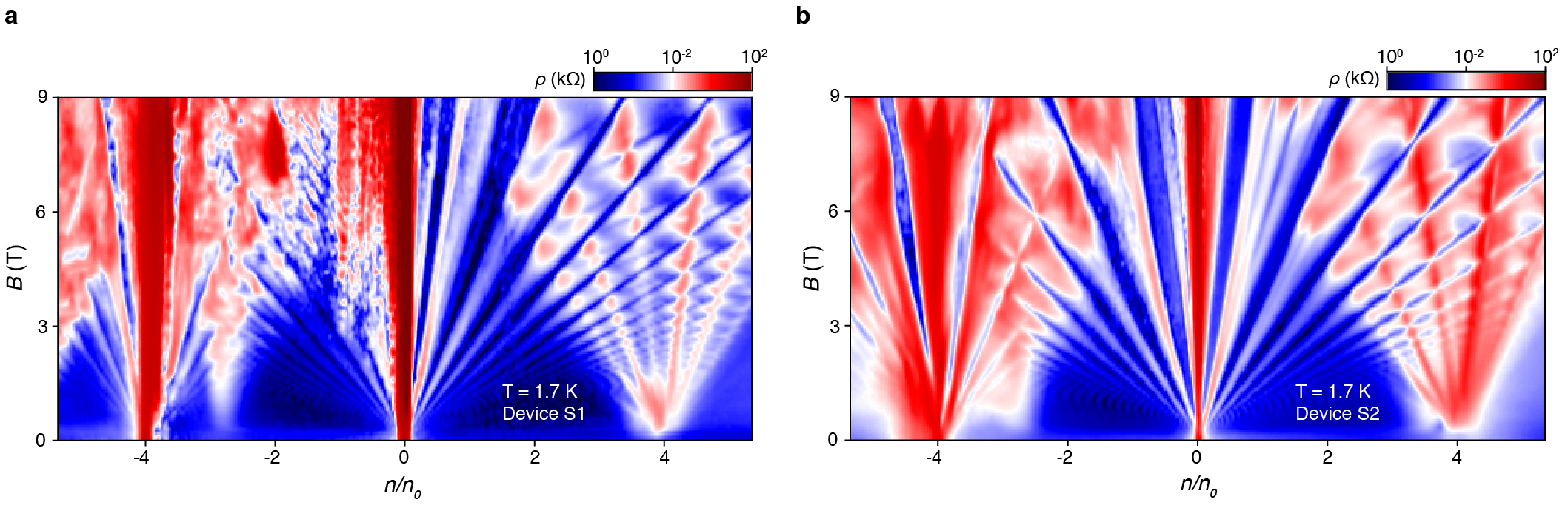} 
\caption{\textbf{Landau fan diagrams for stationary devices.}
Longitudinal resistance versus $B$ for devices \textbf{a,} S1, with $\theta_t \approx \theta_b \approx 0^{\circ}$ and
\textbf{b,} S2, with $\theta_t \approx 60^{\circ}, \theta_b \approx 0^{\circ}.$ Measurements are acquired at $T$ = 1.7 K. $n/n_0=\pm4$ corresponds to full filling of the moir\'e unit cell.
}
\label{fig:S4}
\end{figure*}
%%%%%%%%%%%%%%%%%%%%%%%%%%%%%%%%%%%%%%%%%%%%%%%%%%%%%%%%%%%%%%%%%%%%%%

\begin{center}
\begin{table*}
\centering
\caption{\textbf{Moir\'e wavelength for several top BN configurations in device R1.}}
\label{table:table1}
\vspace{7pt}
\begin{tabular}{|C{75pt}|C{75pt}|C{80pt}|C{80pt}|C{80pt}|C{45pt}|C{45pt} N|}
\hline
$\theta_t$  & n$_{SDP}$ (10$^{12}$ cm$^{-2}$) (doubly-aligned) & n$_{SDP,1}$ (10$^{12}$ cm$^{-2}$) & n$_{SDP,2}$ (10$^{12}$ cm$^{-2}$) & $\lambda$ (nm) (doubly-aligned)& $\lambda_1$ (nm)& $\lambda_2$ (nm)& \\ [10pt]
\hline
unencapsulated & n/a & 2.05 & n/a & n/a & 15.0 & n/a & \\
\hline
-0.7$^{\circ}$ & n/a & 2.05 & 3.22 & n/a & 15.0 & 12.0 & \\
\hline
0$^{\circ}$ & 2.00 & n/a & n/a & 15.2 & n/a & n/a & \\
\hline
9.9$^{\circ}$ & n/a & 2.05 & n/a & n/a & 15.0 & n/a & \\
\hline
59.1$^{\circ}$ & n/a & 2.05 & 3.00 & n/a & 15.0 & 12.4 & \\
\hline
59.4$^{\circ}$ & n/a & 2.05 & 3.72 & n/a & 15.0 & 11.1 & \\
\hline
60$^{\circ}$ & 2.00 & n/a & n/a & 15.2 & n/a & n/a & \\
\hline
\end{tabular}
\end{table*}
\end{center}

\subsection{Determination of moir\'e wavelength and zero top BN angle assignment}

We extract the moir\'e wavelength in our devices via the geometric relation $\lambda^2 = 8/(n_{SDP}\sqrt{3})$. We accurately determine $n_{SDP}$ by projecting the quantum oscillations from the SDP Landau fan to $B=0$. Table~\ref{table:table1} lists $n_{SDP}$ and $\lambda$ for several values of $\theta_t$ in device R1. We additionally measure $n_{SDP}$ in a region of the graphene Hall bar which is not encapsulated by the top BN rotator, such that we are sensitive only to $\theta_b$, and extract a corresponding $\lambda$ = 15.0 nm. This implies nearly perfect alignment of the graphene and bottom BN. The maximum uncertainty in moir\'e wavelength calculated at each configuration is estimated to be $\pm0.1$ nm.

We observe identical $n_{SDP}$ for all $\theta_t$ except at the angles labeled 0$^{\circ}$ and 60$^{\circ}$, in which $n_{SDP}$ becomes marginally smaller. We take the smallest measured value of $n_{SDP}$ to correspond to perfect alignment of the graphene and top BN. Using $\theta_t=0^{\circ}$ and the calculated $\lambda$ = 15.2 nm, we use the geometric relation derived in Ref.~\cite{Yankowitz2012} to extract the lattice mismatch between the graphene and BN to be $\delta\approx$ 1.65\%. We then estimate the misalignment in $\theta_b$ to be less than $\pm$0.15$^{\circ}$, although we are not sensitive to the sign of the misalignment. We use these values to generate the black dashed curve in the inset of Fig.~\ref{fig:4}b in the main text.

We note that if we instead assume $\theta_b\equiv0^{\circ}$, then the corresponding $n_{SDP,1}=2.05\times10^{12}$ cm$^{-2}$ sets a lattice mismatch of $\delta\approx$ 1.67\%. The larger 15.2 nm wavelength computed for the $\theta_t = 0^{\circ}$ and 60$^{\circ}$ configurations (where $n_{SDP}=2.00\times10^{12}$ cm$^{-2}$) could instead be understood to arise owing to a decrease in the lattice mismatch ($\sim$1.3\%), indicating the onset of a commensurate transition, which has not yet been observed in exfoliated graphene--BN heterostructures. However, we are not able to reliably distinguish between these two possibilities within our experimental resolution.

The values of $\theta_b$ and $\theta_t$ are more challenging to precisely determine in the stationary devices owing to the lack of dynamic rotational control. Our method of using fractured BN flakes for both the top and bottom encapsulating layers ensures that the relative twist angle of the two BN layers can be determined optically to better than 1$^{\circ}$. We use the broadened Raman 2D peak to first identify that the graphene is well aligned to both BN layers. However, because stacking sequence of bulk BN is AA' (in which boron and nitrogen atoms alternate between stacked layers)~\cite{Constantinescu2013}, there is ambiguity in whether $\theta_t=0^{\circ}$ or 60$^{\circ}$ in our stationary devices depending on whether the number of layers in the BN flake is even or odd. We therefore rely on the measured band gap --- in particular whether it is enhanced or suppressed from the typical value of a graphene device aligned with a single BN --- to distinguish between $\theta_t=0^{\circ}$ and 60$^{\circ}$.

Figs.~\ref{fig:S4}a-b plot Landau fan diagrams for stationary devices S1 and S2. Graphene is aligned to both BN layers in these devices, with $\theta_t=0^{\circ}$ in device S1 and 60$^{\circ}$ in device S2. Despite the difference in stacking configuration, we observe qualitatively similar features in both fan diagrams, with sequences of quantum oscillations emerging from the PDP and from the SDPs. Although there may be quantitative differences in the Landau level gaps at the SDPs compared to devices in which graphene is aligned to only a single BN layer due to the doubled moir\'e potential strength, we have not investigated these effects in detail in this study.

%%%%%%%%%%%%%%%%%%%%%%%%%%%%%%%%%%%%%%%%%%%%%%%%%%%%%%%%%%%%%%%%%%%%%%
\begin{figure*}[ht]
\includegraphics[width=3.9 in]{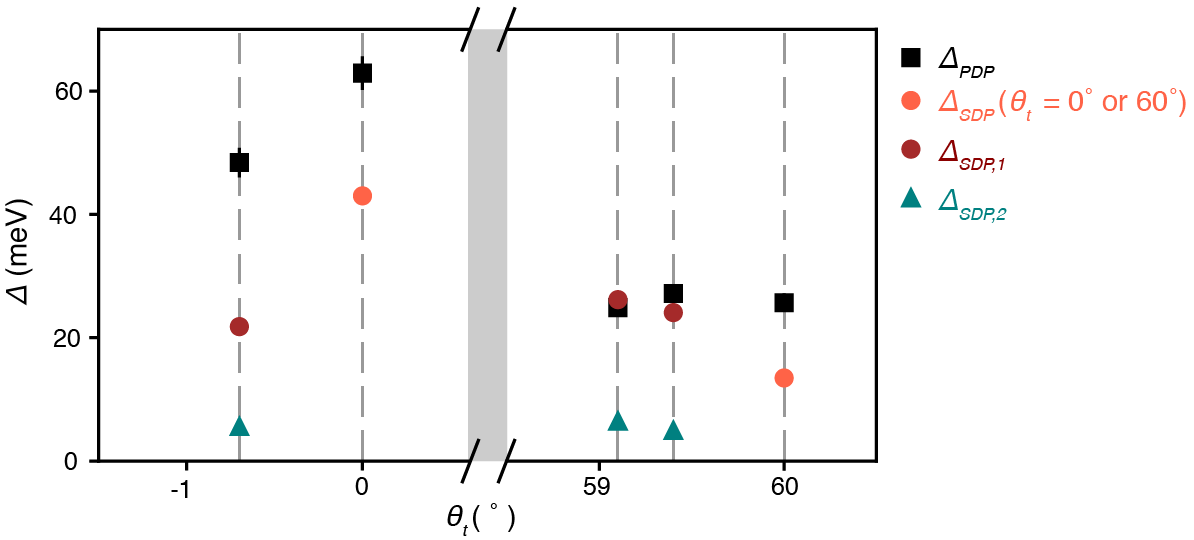} 
\caption{\textbf{Gaps for small top BN twist angles.}
Gaps at the primary Dirac point ($\Delta_{PDP}$, black markers); and secondary Dirac points for the bottom BN/graphene moir\'e pattern ($\Delta_{SDP,1}$, dark red markers), top BN/graphene moir\'e pattern ($\Delta_{SDP,2}$, teal markers). A single SDP is observed for $\theta_t = 0^{\circ}$ and $60^{\circ}$ ($\Delta_{SDP}$, orange markers).
}
\label{fig:S5}
\end{figure*}
%%%%%%%%%%%%%%%%%%%%%%%%%%%%%%%%%%%%%%%%%%%%%%%%%%%%%%%%%%%%%%%%%%%%%%

\subsection{Band gaps for small top BN twist angles}

Fig.~\ref{fig:S5} shows the band gaps measured by thermal activation at the PDP and both valence band SDPs in the case of small, non-zero $\theta_t$. In the case of perfect alignment, only a single SDP gap is observed, as discussed in the main text. We find that the $\Delta_{SDP,2}$ are much smaller than the corresponding $\Delta_{SDP,1}$, however we have not measured these gaps at enough values of $\theta_t$ to conclusively determine a trend with twist angle.

%%%%%%%%%%%%%%%%%%%%%%%%%%%%%%%%%%%%%%%%%%%%%%%%%%%%%%%%%%%%%%%%%%%%%%
\begin{figure*}[ht]
\includegraphics[width=7.0 in]{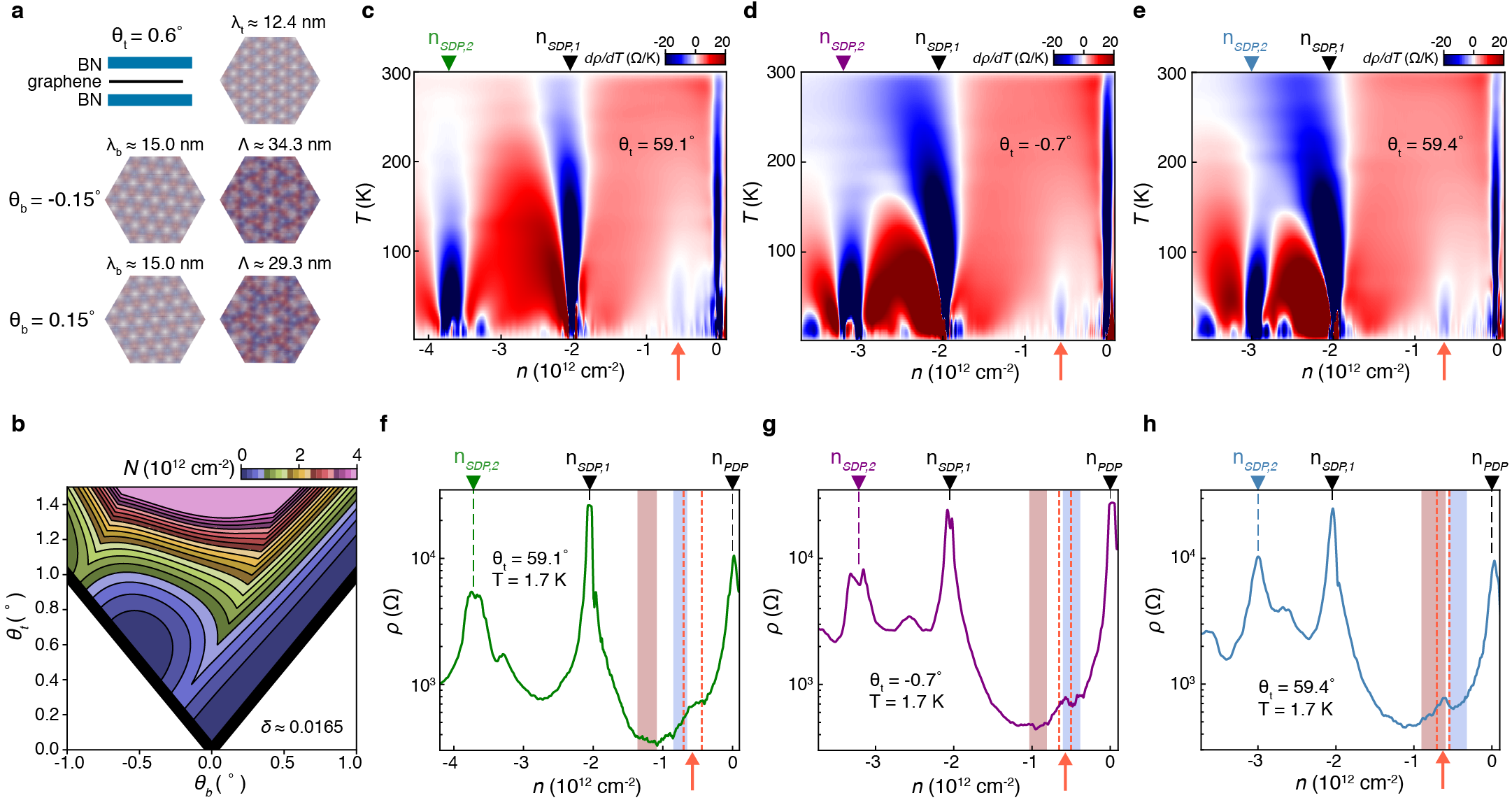}
\caption{\textbf{Second order moir\'e patterns.}
\textbf{a,} Moir\'e patterns for two twist configurations generated from $\theta_b \approx \pm$0.15$^{\circ}$ and $\theta_t \approx$ 0.6$^{\circ}$, labeled with their corresponding first and second order moir\'e wavelengths, $\lambda_{t,b}$ and $\Lambda$.
\textbf{b,} Dispersion map for the density at full filling of the second order moir\'e unit cell, $N$, as a function of $\theta_t$ and $\theta_b$. The lattice mismatch is assumed to be $\delta\approx$1.65\%.
\textbf{c}-\textbf{e,} $d\rho/dT$ as a function of $T$ and $n$ for twist angles $\theta_t$ = 59.1$^{\circ}$, -0.7$^{\circ}$, and 59.4$^{\circ}$. The orange arrows indicate hole carrier densities corresponding to negative values of $d\rho/dT$ that appear near the expected values of $N$ for the given twist configurations.
\textbf{f}-\textbf{h,} $\rho$ as a function of $n$ for twist angles $\theta_t$ = 59.1$^{\circ}$, -0.7$^{\circ}$, and 59.4$^{\circ}$ at $T$ = 1.7 K. The vertical orange dashed lines represent the upper and lower bounds on the negative $d\rho/dT$ regions highlighted in (c)-(e). The range of possible values of $N$ (set by experimental uncertainty) at each value of $\theta_t$ are shown for $\theta_b<0^{\circ}$ (red shaded region) and $\theta_b>0^{\circ}$ (blue shaded region).
}
\label{fig:S6}
\end{figure*}
%%%%%%%%%%%%%%%%%%%%%%%%%%%%%%%%%%%%%%%%%%%%%%%%%%%%%%%%%%%%%%%%%%%%%%

%%%%%%%%%%%%%%%%%%%%%%%%%%%%%%%%%%%%%%%%%%%%%%%%%%%%%%%%%%%%%%%%%%%%%%
\begin{figure*}[ht]
\includegraphics[width=3.25 in]{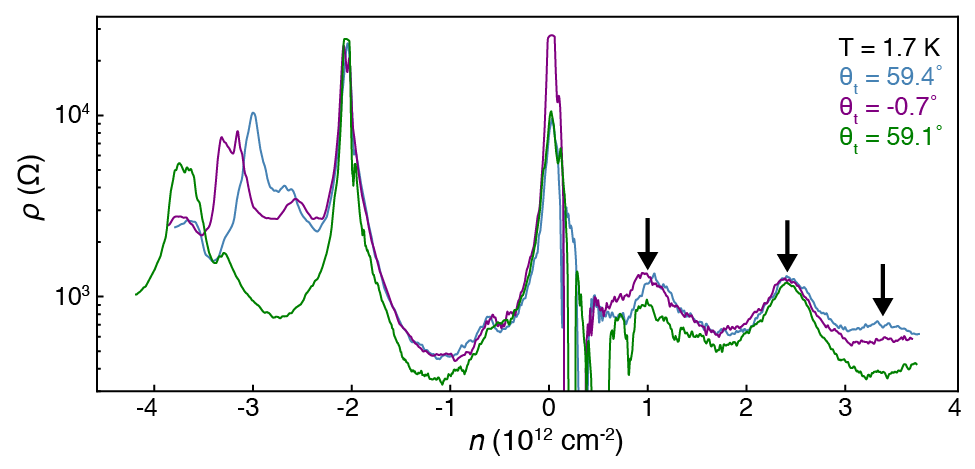} 
\caption{\textbf{Low temperature transport for $\theta_t < 1^{\circ}$.} Three resistance peaks appear for electron-type doping, however their carrier densities do not correspond precisely with their hole-doped counterparts. Regions of apparently vanishing resistance for small electron-type doping are experimental artifacts owing to imperfections in the contacts.}
\label{fig:S7}
\end{figure*}
%%%%%%%%%%%%%%%%%%%%%%%%%%%%%%%%%%%%%%%%%%%%%%%%%%%%%%%%%%%%%%%%%%%%%%

\subsection{Second order moir\'e patterns}

Coexisting moir\'e patterns with different wavelengths may in principle interfere to produce a second-order moir\'e pattern. We determine the wavelength, $\Lambda$, using the dispersion in Ref.~\cite{Yankowitz2012} with lattice constants taken as the moir\'e wavelength of the graphene and bottom BN, $\lambda_b$, and the graphene and top BN, $\lambda_t$. The twist angle between the constituent moir\'e patterns, $\Theta$, is a function of the twist angle of the top and bottom moir\'e patterns relative to the graphene ($\phi_t$ and $\phi_b$, respectively), determined by $\theta_t$, $\theta_b$, and $\delta$~\cite{Yankowitz2012}. Values of $|\phi_t -\phi_b|$, must correctly map to values of $\Theta$ on a 60$^{\circ}$ interval centered at 0$^{\circ}$ for the dispersion to predict $\Lambda$ \cite{Wang2019}. Accounting for this, we have
$$\Lambda = \frac{(1+D)a^2}{\sqrt{2(1+D)(1-cos\Theta) + D^2}}$$
where
$$D = \frac{\lambda_b-\lambda_t}{\lambda_t},\quad \lambda_t \leq \lambda_b$$
$$\tan{\phi_{t,b}} = \frac{-\sin{\theta_{t,b}}}{(1+D)-\cos{\theta_{t,b}}}$$
$$\Theta = \left(|\phi_t-\phi_b|-30^{\circ}\right) \textbf{mod}\:(60^{\circ}) - 30^{\circ}$$

The modular division used here to find $\Theta$ reproduces the output of the piecewise conditional reported in \cite{Wang2019}, with a negligible error produced at the endpoints of the interval.

Fig.~\ref{fig:S6}a shows second order moir\'e patterns for two combinations of top and bottom BN twist angles, with values of $\Lambda$ calculated using the scheme described above. We note that there is an inequivalence in second order moir\'e wavelength for $\theta_b=+$0.15$^{\circ}$ and $\theta_b=-$0.15$^{\circ}$, owing to the difference in moir\'e twist angle $\phi_b$ for each case. Therefore for the top BN twist angle configuration shown ($\theta_t =$ 0.6$^{\circ}$), there are two possible second order moir\'e wavelengths possible if the sign of the bottom BN twist angle is unknown.

Fig.~\ref{fig:S6}b plots the density at full filling of the second order moir\'e unit cell, $N$, determined by $N$ = 8/($\Lambda^2\sqrt{3}$) as a function of $\theta_t$ and $\theta_b$, with an assumed graphene/BN lattice mismatch of $\delta\approx$ 1.65\%. Our transport measurements exhibit dominant resistive peaks corresponding to densities $n_{SDP,1}$ and $n_{SDP,2}$ (Figs.~\ref{fig:4}b-f and ~\ref{fig:S6}c-h) with at least weakly activated behavior and corresponding sequences of quantum oscillations. Furthermore, the measured top BN twist angles corresponding to $n_{SDP,2}$ are consistent with top BN twist angles predicted by the dispersion in Ref.~\cite{Yankowitz2012}, and we observe that $n_{SDP,1}$ is fixed by $\theta_b$ for all values of $\theta_t$. We therefore conduct our analysis of possible second order moir\'e effects on the basis that the densities labeled $n_{SDP,1}$ and $n_{SDP,2}$ correspond to the density at full filling of the bottom and top moir\'e unit cells, respectively.

To compute the expected values of $N$ for both $\theta_b>0^{\circ}$ and $\theta_b<0^{\circ}$, we first take the uncertainty in $\delta$, $n_{SDP,1}$, and $n_{SDP,2}$ into account. We estimate the uncertainty in $n_{SDP,1}$ and $n_{SDP,2}$ to be $\sim\pm0.02 \times 10^{12}$cm$^{-2}$, corresponding to the uncertainty in fitting the trajectories of the quantum oscillations in the Landau fan diagram. Using a standard differential error propagation we estimate the range of values of $\theta_b$ and $\theta_t$ over a range of possible $\delta$ (ranging from $\sim$1.637\% to $\sim$1.654\%). This sets a range of possible values of $N$ at each value of $\theta_t$.

Figs.~\ref{fig:S6}c-e show $d\rho/dT$ as a function of $T$ and $n$ for $\theta_t =$ 59.1$^{\circ}$, -0.7$^{\circ}$ and 59.4$^{\circ}$. We observe regions of negative $d\rho/dT$ (orange arrows, indicating insulating-like behavior) in the vicinity of the expected values of $N$. Figs.~\ref{fig:S6}f-h show $\rho$ as a function of $n$ at $T$ = 1.7 K for the same values of $\theta_t$. The bounds of the negative $d\rho/dT$ regions observed in Figs.~\ref{fig:S6}c-e are marked by orange dashed lines. The computed range of possible values of $N$ at each value of $\theta_t$ are shown for $\theta_b<0^{\circ}$ (red shaded region) and $\theta_b>0^{\circ}$ (blue shaded region). 

We find that there exists some overlap between the computed values of $N$ and the observed regions of negative $d\rho/dT$, suggesting a possible correspondence with second-order moir\'e features. However, for fixed $\delta$ and $n_{SDP,1}$, there appears to be no combination of choices for $n_{SDP,2}$ that generate values of $N$ matching negative $d\rho/dT$ features for all three configurations ($\theta_t=59.1^{\circ}$, -0.7$^{\circ}$, and 59.4$^{\circ}$) simultaneously. Furthermore, there are additional regions of negative $d\rho/dT$ marked by orange arrows in Fig. 4f of the main text at higher densities which remain unexplained. While these may be related to replica features of the low density state around SDPs, their density spacing from SDP features is not identical to $N$.

We additionally plot the transport over a full range of accessible gate voltage in Fig.~\ref{fig:S7}. We observe three features for electron-type doping roughly corresponding to the two SDPs and a lower density feature (black arrows). However, unexpectedly the exact densities of these features do not exactly correspond with their hole-doped counterparts. Therefore, a more detailed study with small $\theta_t$ is necessary to understand all of these extra resistive states in more detail, and in particular how they relate to second order moir\'e patterns. 

%%%%%%%%%%%%%%%%%%%%%%%%%%%%%%%%%%%%%%%%%%%%%%%%%%%%%%%%%%%%%%%%%%%%%%
\begin{figure*}[ht]
\includegraphics[width=7.0 in]{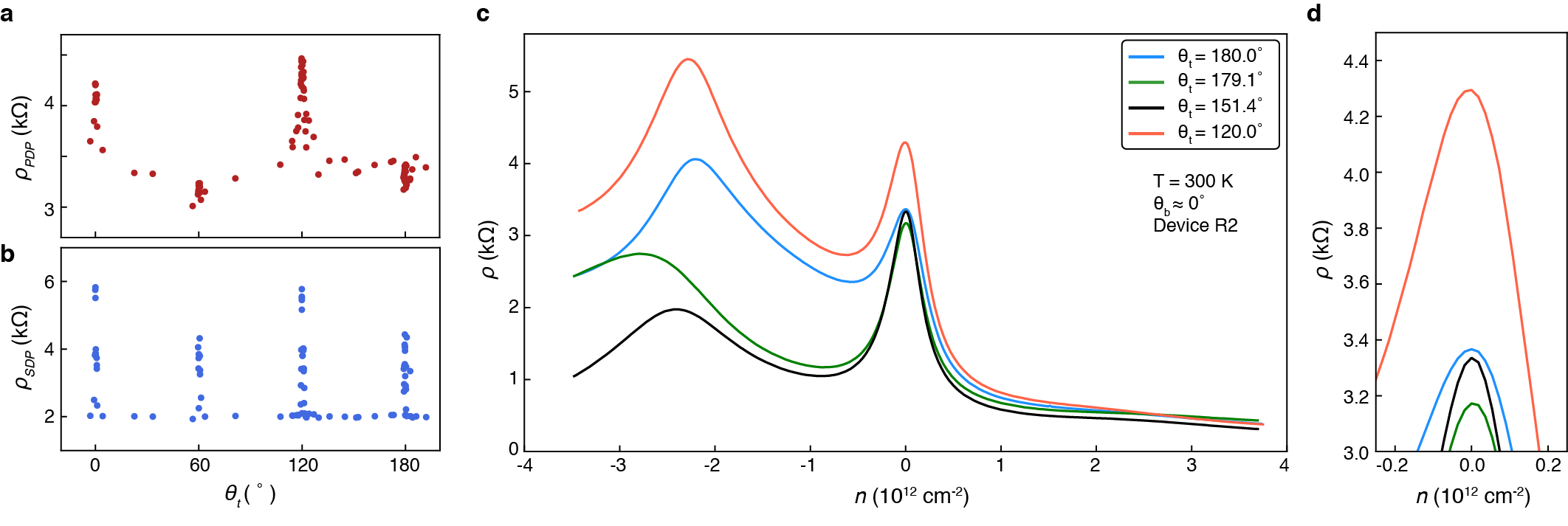} 
\caption{\textbf{Room temperature transport in device R2 ($\theta_b=0\pm0.16^{\circ}$) as a function of $\theta_t$.}
\textbf{a}, $\rho_{PDP}$ and \textbf{b}, $\rho_{SDP}$ at $T=300$ K over a $\sim$200$^{\circ}$ range of $\theta_t$.
\textbf{c,} $\rho$ as a function of $n$ for top BN angles $\theta_t=180^{\circ}$ (blue line), $\theta_t=179.1^{\circ}$ (green line), $\theta_t=151.4^{\circ}$ (black line), and $\theta_t=120^{\circ}$ (orange line).
\textbf{d,} $\rho$ as a function of $n$ near the PDP for the same sweeps shown in \textbf{c}. 
}
\label{fig:S8}
\end{figure*}
%%%%%%%%%%%%%%%%%%%%%%%%%%%%%%%%%%%%%%%%%%%%%%%%%%%%%%%%%%%%%%%%%%%%%%

\subsection{Room temperature transport in device R2}

We measure the room temperature transport of a second rotatable device, R2. Fig.~\ref{fig:S8}a-b shows $\rho_{PDP}$ and $\rho_{SDP}$ as a function of $\theta_t$. Similar to device R1, we observe extrema in both $\rho_{PDP}$ and $\rho_{SDP}$ with 60$^{\circ}$ periodicity. Following the same procedure, we define $\theta_t=0^{\circ}$ as the  angle corresponding to the maximum values observed in both $\rho_{PDP}$ and $\rho_{SDP}$ over a given 120$^{\circ}$ interval. We observe minima in $\rho_{PDP}$ as well as suppressed enhancement of $\rho_{SDP}$ near $\theta_t=60^{\circ}$ in both devices.

Fig.~\ref{fig:S8}c plots gate sweeps at various $\theta_t$ in device R2. Fig.~\ref{fig:S8}d shows a zoomed-in view of the PDP peaks from Fig.~\ref{fig:S8}c. We observe a slight suppression in $\rho_{PDP}$  very near $\theta_t=180^{\circ}$ compared with its value at large misalignment (151.4$^{\circ}$). Comparable behavior is observed in device R1, however at present we do not have a complete model to understand the room-temperature transport response, which is complicated by scattering from acoustic phonons in the graphene and from polar optical phonons in the BN substrate.

\end{document}